\documentstyle[11pt,psfig,amsthm]{article}
\input{amssym.def}
\input{amssym}

\begin{document}
\def\refname{References}

\def\ba{\begin{array}}
\def\ea{\end{array}}

\def\R{{\cal R}^{n+1,2}}
\def\RR{{\cal R}^{n+1,1}}
\def\Rn{{\cal R}^n}

\def\c{\acute{\bf c}}
\def\s{\acute{s}}
\def\a{\acute{\bf a}}
\def\b{\acute{\bf b}}

\def\A{\acute{A}}
\def\B{\acute{B}}
\def\I{\acute{I}}
\def\e{{\bf e}}

\newdimen\LHSskip
\long\def\alignwitheqnos#1#2#3#4{
 \hbox{
    \setbox0=\hbox{$\displaystyle{}#2$}%
     \LHSskip=#1
     \advance\LHSskip by-\wd0
     \rlap{\hskip\LHSskip\unhbox0}
     \hskip #1$\displaystyle{} = #3$
   }
 \vskip-\baselineskip\hfill(#4)%
}

\newdimen\LHSskip
\long\def\alignwithouteqnos#1#2#3{%
 \hbox{
    \setbox0=\hbox{$\displaystyle{}#2$}%
     \LHSskip=#1
     \advance\LHSskip by-\wd0
     \rlap{\hskip\LHSskip\unhbox0}
     \hskip #1$\displaystyle{} = #3$
   }
}

\newbox\bigstrutbox
\setbox\bigstrutbox=\hbox{\vrule height12.5pt depth5pt width0pt}
\def\bigstrut{\relax\ifmmode\copy\bigstrutbox\else\unhcopy\bigstrutbox\fi}

\newbox\Bigstrutbox
\setbox\Bigstrutbox=\hbox{\vrule height17.5pt depth5pt width0pt}
\def\Bigstrut{\relax\ifmmode\copy\Bigstrutbox\else\unhcopy\Bigstrutbox\fi}

\theoremstyle{plain}
\newtheorem{lemma}{Lemma}[section]
\newtheorem{theorem}[lemma]{Theorem}

\def\myeq#1#2{
 \setbox0=\hbox{$\displaystyle #1$}
\vskip.05in \noindent\hskip.5\hsize \hskip-.5\wd0\copy0\hfill(#2)
\vskip.05in}

\renewcommand{\baselinestretch}{1}
\multiply\baselineskip by \baselinestretch

\renewcommand{\theequation}{\arabic{section}.\arabic{equation}}

\def\bt{\begin{theorem}}
\def\et{\end{theorem}}
\def\bc{\begin{corollary}\rm}
\def\ec{\end{corollary}}
\def\ds{\displaystyle}
\def\bl{\begin{lemma}\rm}
\def\el{\end{lemma}}

\def\bi{\begin{itemize}}
\def\ei{\end{itemize}}
\def\bu{\begin{enumerate}}
\def\eu{\end{enumerate}}
\def\bd{\begin{definition}\rm}
\def\ed{\end{definition}}

\def\be{\begin{equation}}
\def\ee{\end{equation}}
\def\no{\noindent}

\def\bx{\begin{proof}}
\def\ex{\end{proof}}

\title{Multisymplectic Geometry and Multisymplectic Preissman Scheme for the KP Equation
\thanks{
Supported by the Special Funds for Major State Basic Research
Projects, G 1999, 032800. }}
\author{Tingting Liu, Menzhao Qin  \\
CCAST(World Laboratory), \\
PoBox 8730, Beijing 100080, China\\
Institute of Computational Mathematics,\\
Academy of Mathematics and Systems Sciences,\\
 Chinese Academy of Sciences,\\
 PoBox 2719, Beijing, 100080, China\\
\texttt{ttliu@lsec.cc.ac.cn, qmz@lsec.cc.ac.cn}}
\date{}
\maketitle
\begin{abstract}
The multisymplectic structure of the KP equation is obtained
directly from the variational principal. Using the covariant De
Donder-Weyl Hamilton function theories, we reformulate the KP
equation to the multisymplectic form which proposed by Bridges.
From the multisymplectic equation, we can derive a multisymplectic
numerical scheme of the KP equation which can be simplified to
multisymplectic forty-five points scheme.
\end{abstract}

\section{Introduction}
\setcounter{equation}{0}
 \label{sec-1} The generalized Kadomtsev-Petviashvili (GKP)
equation is \be (u_t+\partial_x f(u)+u_{xxx})_x+\sigma u_{yy}=0 \quad (t>0,
-\infty<x,y<\infty)\label{1.1}
 \ee where $\sigma$ is a constant, $f(u)$ is some smooth function.
 The usually KP equation occurs for $f(u)$ quadratic and it is regarded as a
two-dimensional generalization of the Korteweg-de Vries (KdV)
equation. It describes the evolution of long water waves of small
amplitude if they are weakly two-dimensional. In the case of
$f(u)=3u^2$ and $\sigma=-3$, Equation (\ref{1.1}) is usually
called the KPI equation, whereas the KPII equation with
$f(u)=3u^2$ and $\sigma=3$. As a soliton equation important from
analytical and numerical point of view, the KP equation is one of
the few known completely integrable equations in the
multi-dimensional soliton equations. Thus, in the last few years,
considerable attention has been paid to the KP equation. Although
considerable interest has been focused on the KP equation, the
numerical scheme analysis literature for the KP equation is
extremely small. As far as we are aware of, Katsis proposed
explicit finite difference method [15], the results of evolution
of lump solution for the KP equation was given by A.A. Minzoni
[17], X.P. Wang et. studied the instability of a generalized KP
equation [16], B.F. Feng and T. Mitsui took the linearized
implicit method to the KP equation [13].

In this paper, we try describe KP equation in the language of
multisymplectic geometry. Recently, for first order field theory,
i.e., the Lagrangian density depends on the state variables and
their first order derivatives, Marsden, Patrix and Shkoller [7]
derived numerical methods for the first order field theories.
However the lagrangian density of the KP equation is not
first-order, therefore MPS theory can not be applied directly. In
[10], the authors focus their attention on the KdV equation whose
Langrangian density is second order. The langrangian density of
the KP equation is truly third-order. In this paper, we give the
multisymplectic structure of the KP equation directly from the
variational principal. In [5], the author proposed the Cartan form
is not necessarily unique, we find it was caused by higher-order
mixed multiple integral in using Stokes' formula in actually
calculus variation.

In Lagrange mechanics, we know the Euler-Lagrange equation can be
write as
\begin{equation}
\frac{d}{dt}\frac{\partial L}{\partial \dot{q}^i}-\frac{\partial
L}{\partial q^i}=0
\end{equation}
Taking the Legandre transform of Lagrange density
$L:p^i=\frac{\partial L}{\partial \dot{q}^i}$, we can rewrite
equation (1.2) as regular Hamilton equation
\begin{equation}
\left\{
\begin{array}{l}
\frac{dp^i}{dt}=-\frac{\partial H}{\partial q^i} \\
\frac{dq^i}{dt}=\frac{\partial H}{\partial p^i}
\end{array}
\right.
\end{equation}
where $H=p^i\dot{q}^i-L(q^i,\dot{q}^i,t)$. With the covariant De
Donder-Weyl Hamilton function theories [18], we can reformulate
the partial differential equation to the following form
\begin{equation}
\left\{
\begin{array}{l}
\frac{\partial H}{\partial \pi_i^{\mu}}=\partial_{\mu}q^i \\
\frac{\partial H}{\partial q^i}=-\partial_{\mu}\pi_i^{\mu}
\end{array}
\right.
\end{equation}
where $ \pi_i^{\mu}=\frac{\partial L}{\partial\partial_{\mu}q^i}
$. According to this methods, we can rewrite KP equation to the
multisymplectic form that introduced by Bridges [3].

Multisymplectic equation have the important multisymplectic
conservation laws. In the numerical study, we also hope that the
numerical approximations can preserve the multisymplectic
conservation laws. Similar to the method [9], we show that the
Preissman scheme is a multisymplectic scheme for the KP equation.
Though the Preissman  scheme is multisymplectic, it need more
computational memory, so we reduce it to a multisymplectic
forty-five points scheme. Using the forty-five points scheme, we
get some numerical results on soliton and solitary waves over long
time intervals.

In section 2, we describe the multisymplectic geometry of the KP
equation entirely in the framework of the variational principal.
Section 3 is devoted to the analysis of multisymplectic Preissman
scheme and reduce it to a multisymplectic forty-five points scheme
. In section 4, some numerical results on soliton and solitary
waves over long time intervals be given.

\section{Multisymplectic Geometry of the KP Equation}
\setcounter{equation}{0}

We now review some aspects of multisymplectic geometry.

Let $X$ be an orientable $(n+1)$-dimensional parameter space (
which is usually space time ) and let $\pi_{XY}:Y\rightarrow X$ be
a fiber over $X$. Section $\varphi :X\rightarrow Y$ of this
covariant configuration bundle is the physical fields. Coordinates
on $X$ are denoted by $x^\mu,\mu=1,2,\cdots,n,0$. In general,
$x^0$ denotes the time coordinate. The parameter $n$ denotes the
number of spatial variables. In this paper, we just discuss the
case of $n>0$. Adapted coordinates on $Y$ are $y^A$ along the
fibers $Y_{x}=\pi_{XY}^{-1}(x),x\in X$, $A=1,2,\cdots,N$. $N$
denote the fibers dimensions. Consider a $k^{th}$ order lagrangian
density $L$, viewed as a fiber-preserving map $L:
J^kY\rightarrow\land^{n+1}X$. $J^k(Y)$ denote the $k^{th}$-order
jet bundle over $Y$ which can be induced by $J^1(\cdots
(J^1(Y)))$. We let $T_{x}X$ denote the tangent space of $X$ at
$x$. Denote the derivative of the map $\pi_{XY}$ in the direction
$w$ by $T_{\pi_{XY}}\cdot w$.

At first, we introduce the first jet bundle.
\newline
{\bf Definition 2.1 } The first jet bundle over $Y$ is a fiber bundle denoted
by $J^1(Y)$ whose fiber over $y\in Y_x=\pi_{XY}^{-1}(x),x\in X$ consists of those linear
mappings $\gamma:T_xX\rightarrow T_yY$ satisfying
$$
T_{\pi_{XY}}\circ \gamma =Identity.
$$
If $\varphi:X\rightarrow Y$ is a section of $Y$, $j^1(\varphi)$ is a section of $J^1(Y)$ and in coordinates, $j^1(\varphi)$ is given by
$$
(x^\mu,\varphi^A(x^\mu),\partial_\mu\varphi^A(x^\mu)),\quad \mu=1,2,\cdots , n,0,
$$

Similarly, higher order jet bundle $J^m(Y)$ is defined by $J^1(J^{m-1}(Y))$.
\newline
{\bf Definition 2.2 } The $k^{th}$-order jet bundle over $Y$ is a fiber bundle denoted
by $J^k(Y)$ whose fiber over $\gamma \in J^{k-1}(Y)_{y},y\in Y$ consists of those linear
mappings $s:T_xX\rightarrow T_\gamma J^{k-1}(Y)$ satisfying
$$
T_{\pi_{X,J^{k-1}Y}}\circ s =Identity.
$$
We let $j^k(\varphi)=j^1(\cdots(j^1(\varphi)):x\rightarrow T_xj^{k-1}(\varphi)$
denote $k^{th}$-order jet prolongation of the section $\varphi :X\rightarrow Y$,
in which $j^{k-1}(\varphi)$ is a section of jet bundle $J^{k-1}(Y)$.
Thus, $j^k(\varphi)$ is given in coordinates
$$
(x^\mu,\varphi^A(x^\mu),\partial_\mu\varphi^A(x^\mu),\cdots, \partial_{\mu_n}\partial_{\mu_{n-1}}\cdots \partial_{\mu_0}\varphi^A(x^\mu)),
$$

Given a $k^{th}$-order lagrangian density $L:J^kY\rightarrow\wedge^{n+1}x$, the basis geometric
object in the classical calculus of variations is the $(n+1)$-form $\theta_L$ on $J^{2k-1}(Y)$, which was called the Cartan form.

 The KP equation can be written as
\begin{equation}
(2u_t+6uu_x+u_{xxx})_x+ \sigma u_{yy}=0.
\end{equation}
The 2 multiplying $u_t$ is added for notational convenience, it
can be eliminated by scaling $t$. In this paper, we consider the
KPI equation.

To put the KP equation in the variational frame work, we let
$\varphi_{xx}=u$, then $\varphi$ satisfies equation:
\begin{equation}
2\varphi_{xxxt} +
6\varphi_{xx}\varphi_{xxxx}+6\varphi_{xxx}^2+\varphi_{xxxxxx}+\sigma
\varphi_{xxyy}=0.
\end{equation}
The search for a variational principle is equivalent to the
inverse problem of the calculus of variations, i.e., the existence
and formulation of variational principles for systems of nonlinear
partial differential equations. The existence of a variational
principle for a differential equation is equivalent to determining
whether or not an operator is a potential operator. According to
Vainberg theorem [2], in order that operator $N$ be potential
operator, to summarize, it is necessary and sufficient that the
Gateau derivative of the operator $N$ is symmetry. The theorem
stated in terms of the Gateau derivative, but we assume that the
Frechet derivative exists in application. Let $N$ is an operator
which defined in an appropriate function space $E$ (typically $E$
is Banach space), then $N$ be potential operator if
 $N'_u=$ {\it \~N}$'_u$, where
$$N'_u\varphi=\lim _{\varepsilon\rightarrow 0}\frac{N(u+\varepsilon\varphi)-N(u)}{\varepsilon}=[\frac{\partial}{\partial\varepsilon}N(u+\varepsilon\varphi)]\mid_{\varepsilon =0},$$
and {\it \~N}$'_u$ is the adjoint operator of $N'_u$. If the operator is a potential operator, the potential $F$ is given by
$$
F=\int u\int_0^1 N(\lambda u) d\lambda dV.
$$
where $\int dV$ represents integration over the physical domain
and $\int_0^1d \lambda$ represents integration over the scalar
variable $\lambda$.
We test the operator
$$N(\varphi)=2\varphi_{xxxt} +
6\varphi_{xx}\varphi_{xxxx}+6\varphi_{xxx}^2+\varphi_{xxxxxx}+\sigma
\varphi_{xxyy}
$$
and find $N'_\varphi=$ {\it \~N}$'_\varphi$. Hence,
$$
\begin{array}{rcl}
F(\varphi)&=&\int\varphi\int_0^1 N(\lambda\varphi)d\lambda dV \\
&=&\int \varphi\int ^1_0[2\lambda\varphi_{xxxt} +
6\lambda^2\varphi_{xx}\varphi_{xxxx}+6\lambda^2\varphi_{xxx}^2+\lambda\varphi_{xxxxxx}+\sigma\lambda
\varphi_{xxyy}]d\lambda dV\\
&=&\int \varphi(\varphi_{xxxt}+2\varphi_{xxx}^2+2\varphi_{xx}\varphi_{xxxx}+\frac{1}{2}\varphi_{xxxxxx}+\frac{\sigma}{2}\varphi_{xxyy})dV
\end{array}
$$
To obtain $F(u)$ in a more familiar form, integrate by part and
discard the integration over the boundary since it has nothing
with the Lagrangian density, then we get potential
$$
F(\varphi)=\int(\varphi_{xx}\varphi_{xt}-\frac{1}{2}\varphi_{xxx}^2+\frac{\sigma}{2}\varphi_{xy}^2+\varphi_{xx}^3)dV.
$$
So we can determined that the Lagrangian density for
equation (2.2) is
$$
L(j^3(\varphi))=(\varphi_{xx}\varphi_{xt}-\frac{1}{2}\varphi^2_{xxx}+\frac{\sigma}{2}\varphi^2_{xy}+\varphi^3_{xx})dx\wedge dy\wedge dt.
$$
Corresponding to Lagrangian density $L(j^3(\varphi))$, the action
function is defined as following:
$$
S(\varphi)=\int_U L(j^3(\varphi)),\quad U \hbox{ is a open set of } X.
$$

Let G be the Lie group of $\pi_{XY}$ bundle automorphisms $\eta_Y$
covering $\eta _X$. Denote $\eta _Y^{\lambda }$ an smooth path in
G such that
$$
\overline \varphi=\eta^\lambda_Y \circ \varphi \circ
(\eta^\lambda_X)^{-1}.
$$
The vector field of $\eta_Y^\lambda$ is
$$
V=\frac {d}{d\lambda}\mid _{\lambda=0} \overline \varphi =\left[
\begin{array}{cccc}
V^x\\
V^y\\
V^t\\
V^\varphi\\
\end{array}
\right].
$$

We say that $\varphi$ is a extremum of $S$ is
$$
\frac{d}{d\lambda}\mid _{\lambda=0} S(\overline \varphi)=0.
$$
Now we consider the variation
$$
\frac{d}{d\lambda}\mid _{\lambda=0} S(\overline
\varphi)=\frac{d}{d\lambda}\mid _{\lambda=0} \int_{\eta_X^\lambda
U}{(\overline \varphi_{\overline x \overline x} \overline
\varphi_{\overline x \overline t} - \frac{1}{2} \overline
\varphi^2_{\overline x \overline x \overline x}+\frac{\sigma}{2} \overline
\varphi^2_{\overline x \overline y}+\overline \varphi^3_{\overline
x \overline x}) d \overline x\wedge d \overline y \wedge d
\overline t}
$$
where
$$
\eta_X^\lambda \left[
\begin{array}{cccc}
x\\
y\\
t
\end{array}
\right] = \left[
\begin{array}{cccc}
\overline x\\
\overline y\\
\overline t
\end{array}
\right].
$$

A direct computation shows
$$
\frac d{d\lambda} \mid _{\lambda=0} S(\overline \varphi)=I_1+I_2,
$$
in which
\begin{equation}
\begin{array}{rcl}
I_{1} & =&
\int_{U}-(2\varphi_{xxxt}+6\varphi_{xx}\varphi_{xxxx}+6\varphi^2_{xxx}+\varphi_{xxxxxx}+
\sigma\varphi_{xxyy}) \\
& & (V^\varphi-\varphi_xV^x-\varphi_yV^y-\varphi_tV^t)dx\wedge dy
\wedge dt,
\end{array}
\end{equation}

\begin{equation}
\begin{array}{rcl}
I_2 & = & \int_{\partial U
}[(\varphi_{xx}\varphi_{xt}-\frac{1}{2}\varphi^2_{xxx}+\frac{\sigma}{2}
\varphi^2_{xy}+\varphi^3_{xx})V^t \\
& & +\varphi_{xx}(V_x^\varphi-\varphi_xV^x_x-\varphi_yV^y_x-\varphi_tV^t_x-
\varphi_{xx}V^x-\varphi_{xy}V^y-\varphi_{xt}V^t)]dx\wedge dy \\
& &[-(\varphi_{xx}\varphi_{xt}-\frac{1}{2}\varphi^2_{xxx}+\frac{\sigma}{2}\varphi_{xy}^2+\varphi_{xx}^3)V^y
\\
 & & +\sigma\varphi_{xy}(V^\varphi_x-\varphi_x
V^x_x-\varphi_yV^y_x-\varphi_tV^t_x-\varphi_{xx}V^x-\varphi_{xy}V^y
-\varphi_{xt}V^t)]dx\wedge
dt\\
& &+[(\varphi_{xx}\varphi_{xt}-\frac{1}{2}\varphi^2_{xxx}+\frac{\sigma}{2}\varphi_{xy}^2+\varphi_{xx}^3)V^x\\
& &
+(\varphi_{xt}+3\varphi_{xx}^2+\varphi_{xxxx})(V^\varphi_x-\varphi_xV^x_x-\varphi_yV^y_x-\varphi_tV^t_x-\varphi_{xx}V^x
\\
& & -\varphi_{xy}V^y-\varphi_{xt}V^t)-\varphi_{xxx}(V^\varphi_{xx}-\varphi_{x}V^x_{xx}-2\varphi_{xx}V_x^x-\varphi_yV_{xx}^y \\
& & -2\varphi_{xy}V^y_x-\varphi_tV^t_{xx}-2\varphi_{xt}V^t_x-\varphi_{xxy}V^y-\varphi_{xxt}V^t-\varphi_{xxx}V^x)\\
& & +
(-2\varphi_{xxt}-6\varphi_{xx}\varphi_{xxx}-\sigma\varphi_{xyy}-\varphi_{xxxxx})(V^\varphi-\varphi_xV^x
-\varphi_yV^y-\varphi_tV^t)]dy\wedge dt.
\end{array}
\end{equation}

In[5], the author introduced the Lepagean equivalents which is a
generalization of the Poincare-Cartan form and proposed the Carton
form is not necessarily unique for higher order. The author also
pointed out that every Lagrangian density has a Lepagean
equivalent on $ J^{2k-1}Y$. It turns out that Lepagean equivalents
exists on jet bundles of order $2k-1$ or higher, but not
necessarily on jet bundle of lower order. The author presented
that: As being the principal part of a Lepagean equivalents, the
Cartan form always exist and are typically nonunique unless $k=1$.
we find it is caused by the higher-order mixed multiple integral
in using Stokes' formula in actually calculus of variations. By
$I_2$, we can define a Cartan form
\begin{equation}
\begin{array}{rcl}
\theta_L& = & (\frac
{1}{2}\varphi^2_{xxx}-\frac{\sigma}{2}\varphi^2_{xy}+2\varphi_x\varphi_{xxt}+6\varphi_x\varphi_{xx}\varphi_{xxx}+
\sigma\varphi_x\varphi_{xyy}+\varphi_x\varphi_{xxxxx}\\
& &
-\varphi_{xx}\varphi_{xt}-2\varphi^3_{xx}-\varphi_{xx}\varphi_{xxxx})dx\wedge
dy\wedge dt\\
& &
+(-2\varphi_{xxt}-6\varphi_{xx}\varphi_{xxx}-\sigma\varphi_{xyy}-\varphi_{xxxxx})d\varphi
\wedge dy\wedge dt \\
& & +(\varphi_{xt}+3\varphi^2_{xx}+\varphi_{xxxx})d\varphi_{x}\wedge
dy\wedge dt-\varphi_{xxx}d\varphi_{xx}\wedge dy\wedge dt\\
& & -\sigma \varphi_{xy}d\varphi_x\wedge dx\wedge dt +
\varphi_{xx}d\varphi_x\wedge dx\wedge dy.
\end{array}
\end{equation}
Since
$$
\begin{array}{l}
j^5(\varphi)^{\ast }dx=dx, \quad j^5(\varphi)^{\ast}dy=dy,\quad j^5(\varphi)^{\ast }dt=dt \\
j^5(\varphi)^{\ast }d\varphi=
\varphi_xdx+\varphi_ydy+\varphi_tdt,\\
j^5(\varphi)^{\ast }d\varphi_x=
\varphi_{xx}dx+\varphi_{xy}dy+\varphi_{xt}dt,\\
j^5(\varphi)^{\ast }d\varphi_{xx}=
\varphi_{xxx}dx+\varphi_{xxy}dy+\varphi_{xxt}dt,
\end{array}
$$
we have
$$
I_2=\int_{\partial U}{j^5(\varphi)^{\ast }(j^5(V)\rfloor \theta
_L)}
$$

here $j^5(V)$ is the jet prolongation of the vector field $V$ [8].
The multisymplectic form is the 4-form $\Omega _L=-d\theta _L$.
Form $\theta _L$ defines a multisymplectic structure on jet bundle
$J^5(Y)$.

Now, we consider the Euler-Lagrange equation for the action
function $S(\varphi)$.

Since $L(j^3(\overline{\varphi}))=j^5(\overline \varphi)^{\ast }\theta _L$,
we have
$$
\begin{array}{rcl}
\frac d{d\lambda }\mid _{\lambda=0}\int_{\eta_X^\lambda
U}{L(j^3(\overline \varphi))} & = & \frac d {d\lambda
}\mid_{\lambda=0}\int_{\eta ^\lambda _X
U}{j^5(\overline\varphi)^{\ast }\theta _L}\\
& = & \frac d{d\lambda}\mid_{\lambda=0}\int_{\eta_X^\lambda
U}{j^5(\eta _Y^\lambda \circ \varphi \circ
(\eta_X^\lambda)^{-1})^{\ast }\theta_L}\\
& = & \frac d{d\lambda}\mid_{\lambda=0}\int_{\eta_X^\lambda
U}{((\eta_X^\lambda)^{-1})^{\ast }j^5(\varphi)^{\ast
}j^5(\eta^\lambda_{Y})^{\ast}\theta_L}\\
& = & \frac d{d\lambda}\mid_{\lambda=0}\int_{U}{j^5(\varphi)^\ast
j^5(\eta_Y^\lambda)^\ast\theta_L}\\
& = & \int_{U}{j^5(\varphi)^\ast \pounds _{j^5(V)}\theta_L}
\end{array}
$$
in which symbol $\pounds$ denotes the Lie derivative.

By the Cartan's Magic formula [6]
$$
\pounds_{j^5(V)}\theta_L = -
{j^5(V)}\rfloor\Omega_L+d(j^{5}(V)\rfloor\theta_L)
$$
We can obtain
$$
\frac
d{d\lambda}\mid_{\lambda=0}S(\overline\varphi)=-\int_U{j^5(\varphi)^\ast(j^5(V)\rfloor\Omega_L)}+\int_{\partial
U}{j^5(\varphi)^\ast(j^5(V)\rfloor\theta_L)}
$$
If $V$ is a vector field with compact support, we have
$$
\int_{\partial U}{j^5(\varphi)^{\ast }(j^5(V)\rfloor \theta_L)=0}
$$
Hence, a necessary condition for $\varphi$ to be an extremum is
that
$$
\int_U{j^5(\varphi)^{\ast}(j^5(V)\rfloor\Omega_L)}=0
$$ for any $V$ with compact support. By compute the integral and
obtain that
\begin{equation}
\begin{array}{rcl}
j^5(\varphi)^\ast
(j^5(V)\rfloor\Omega_L) & = & (2\varphi_{xxxt}+6\varphi_{xx}\varphi_{xxxx}+6\varphi_{xxx}^2+\varphi_{xxxxxx}\\
& &
+\sigma\varphi_{xxyy})(V^{\varphi}-\varphi_xV^x-\varphi_yV^y-\varphi_tV^t)
\end{array}
\end{equation}
Remark: We can get the Euler-lagrange equation from the vertical
variation $V^\varphi$, and the $V^x$ and $V^y$ directions
horizontal variations gives the law of conservation of momentum.
The law of conservation of energy can be obtained along
time-direction horizontal variation.

Taking the $\pi_{XY}$-vertical vector field $V$ and using the
standard method from the calculus of variations, we obtain that
$\varphi$ satisfies
\begin{equation}
2\varphi_{xxxt}+6\varphi_{xx}\varphi_{xxxx}+6\varphi^2_{xxx}+\varphi_{xxxxxx}+\sigma\varphi_{xxyy}=0
\end{equation}
i.e., the equation (2.2). So, for any vector field $V$,
\begin{equation}
j^5(\varphi)^\ast(j^5(V)\rfloor\Omega_L)=0
\end{equation}
holds. A short computation verifies that
\begin{equation}
j^5(\varphi)^\ast(P\rfloor\Omega_L)=0
\end{equation}
where $P\in TJ^5(Y)$ and is $T_{\pi_{Y,J^5(Y)}}$-vertical. For
any $W\in TJ^5(Y)$, there exists vector field $V$, such that
\begin{equation}
W=j^5(V)+P
\end{equation}
So, by (2.8)-(2.10), if $\varphi$ is an extremum of $S$,
$j^5(\varphi)^\ast(W\rfloor\Omega_L)$ mush vanish for any vector
field $W\in TJ^5(Y)$, thus, we get the Euler-lagrange equation
\begin{equation}
j^5(\varphi)^\ast(W\rfloor\Omega_L)=0
\end{equation}
for any vector field $W\in TJ^5(Y)$. In the following part, we
consider the multisymplectic form formula and a corollary of the
multisymplectic form formula. About the multisymplectic form
formula for first order field theories, please refer to [11].
\newline
{\bf Theorem 2.3} let $\eta^\lambda_Y$ and $\xi_Y^\lambda$ demote
 two one-parameter symmetry groups of equation (2.11) and the
 corresponding vector fields are $V$ and $W$. Then we have the
 multisymplectic form formula
 \begin{equation}
 \int_{\partial
U}{j^5(\varphi)^\ast(j^5(V)\rfloor j^5(W)\rfloor\Omega_L)}=0.
 \end{equation}
Proof: Since $j^5[W,V]=[j^5(W),j^5(V)]$, from (2.11) we have
\begin{equation}
\begin{array}{rcl}
0& = & \int_U{j^5(\varphi)^\ast(j^5[W,V]\rfloor\Omega_L)} \\
& = & \int_U{j^5(\varphi)^\ast([j^5(W),j^5(V)]\rfloor\Omega_L)}\\
& = &
\int_U{j^5(\varphi)^\ast(\pounds_{j^5(W)}(j^5(V)\rfloor\Omega_L)-j^5(V)\rfloor\pounds_{j^5(W)}\Omega_L)}.
\end{array}
\end{equation}
Because $\eta_Y^\lambda$ and $\xi_Y^\lambda$ are two one-parameter
symmetry groups of equation (2.11), so for any vector field $Q\in
TJ^5(Y)$, we have
\begin{equation}
\frac
d{d\lambda}\mid_{\lambda=0}j^5(\eta^\lambda_Y\circ\varphi\circ(\eta_X^\lambda)^{-1})^\ast(Q\rfloor\Omega_L)=j^5(\varphi)^\ast\pounds_{j^5(V)}(Q\rfloor\Omega_L)=0,
\end{equation}
\begin{equation}
\frac
d{d\lambda}\mid_{\lambda=0}j^5(\xi^\lambda_Y\circ\varphi\circ(\xi_X^\lambda)^{-1})^\ast(Q\rfloor\Omega_L)=j^5(\varphi)^\ast\pounds_{j^5(W)}(Q\rfloor\Omega_L)=0.
\end{equation}
Thus (2.13) becomes
\begin{equation}
\begin{array}{rcl}
0& = & -\int_U{j^5(\varphi)^\ast(j^5(V)\rfloor\pounds_{j^5(W)}\Omega_L)} \\
& = & -\int_U{j^5(\varphi)^\ast(j^5(V)\rfloor d(j^5(W)\rfloor\Omega_L))} \\
& = & \int_U{j^5(\varphi)^\ast(j^5(V)\rfloor
d(\pounds_{j^5(W)}\theta _L))}.
\end{array}
\end{equation}
$j^5(V)\rfloor j^5(W)\rfloor\Omega_L$ can be written as
\begin{equation}
\begin{array}{rcl}
j^5(V)\rfloor j^5(W)\rfloor\Omega_L & = & j^5(V)\rfloor d(j^5(W)\rfloor\theta_L)-j^5(V)\rfloor \pounds_{j^5(W)}\theta_L \\
& = & \pounds_{j^5(V)}( j^5(W)\rfloor\theta_L)-j^5(V)\rfloor
\pounds_{j^5(W)}\theta_L- d(j^5(V)\rfloor j^5(W)\rfloor\theta_L).
\end{array}
\end{equation}

So from stokes' formula we can obtain that
\begin{equation}
\begin{array}{rcl}
& & \int_{\partial U}{j^5(\varphi)^\ast(j^5(V)\rfloor j^5(W)\rfloor\Omega_L)} \\
& = & \int_{\partial U}{j^5(\varphi)^\ast(\pounds_{j^5(V)}(j^5(W)\rfloor\theta_L)-j^5(V)\rfloor\pounds_{j^5(W)}\theta_L-d(j^5(V)\rfloor j^5(W)\rfloor \theta_L))} \\
& = & \int_{U}{j^5(\varphi)^\ast d
(\pounds_{j^5(V)}(j^5(W)\rfloor\theta_L)-j^5(V)\rfloor\pounds_{j^5(W)}\theta_L)}
\\
& = &  \int_{U}{j^5(\varphi)^\ast
(\pounds_{j^5(V)}\pounds_{j^5(W)}\theta_L+\pounds_{j^5(V)}(j^5(W)\rfloor\Omega_L)-d({j^5(V)}\rfloor
\pounds_{j^5(W)}\theta_L))}.
\end{array}
\end{equation}
By
\begin{equation}
\begin{array}{rcl}
\pounds_{j^5(V)}\pounds_{j^5(W)}\theta_L & = & j^5(V)\rfloor
d(\pounds_{j^5(W)}\theta_L)+d(j^5(V)\rfloor\pounds_{j^5(W)}\theta_L),
\end{array}
\end{equation}
we have
\begin{equation}
\begin{array}{rcl}
& & \int_{\partial U}{j^5(\varphi)^\ast(j^5(V)\rfloor j^5(W)\rfloor\Omega_L)} \\
& = & \int_{U}{j^5(\varphi)^\ast(j^5(V)\rfloor
d(\pounds_{j^5(W)}\theta_L)+\pounds_{j^5(V)}(j^5(W)\rfloor\Omega_L))}.
\end{array}
\end{equation}

Hence, by (2.14) and (2.16), we obtain
$$
\int_{\partial U}{j^5(\varphi)^\ast(j^5(V)\rfloor
j^5(W)\rfloor\Omega_L)}=0.
$$

Although the covariant Legendre transformation ( or complete
Legendre transformation ) which transform time and space variable
simultaneously are not necessarily unique, for this fixed Cartan
form $\theta_L$, we can construct corresponding covariant
Lengendre transformation of Lagrangian density $L$. Let
$v=\varphi_x, u=\varphi_{xx}, w=\varphi_{xy}, p=\varphi_{xt}$,
taking the covariant Legendre transform of Lagrangian density $L$:
$$
\begin{array}{rcl}
&  & p^x=-2\varphi_{xxt}-6\varphi_{xx}\varphi_{xxx}-\sigma\varphi_{xyy}-\varphi_{xxxxx}, \\
&  & p^{xx}= \varphi_{xt}+3\varphi_{xx}^2+\varphi_{xxxx},\quad  p^{xt}=\varphi_{xx},\\
&  & p^{xy}=\sigma\varphi_{xy}, \quad p^{xxx}=-\varphi_{xxx},
\end{array}
$$
According to the covariant De Donder-Weyl Hamilton function
theories [18] and the multisymplectic concept introduced by
Bridges [3], KP equation can be reformulated as a system of ten
first-order partial differential equations which can be written in
the form:
\begin{equation}
\begin{array}{rcl}
M\Bbb{Z}_t+K\Bbb{Z}_x+L\Bbb{Z}_y=\nabla S(\Bbb{Z}),\\
\Bbb{Z}=(\varphi,v,u,w,p,p^x,p^{xx},p^{xy},p^{xt},p^{xxx})^T\in\Bbb{R}^{10},
\end{array}
\end{equation}
where
$$
M=\left(
\begin{array}{cccccccccc}
    0 & 0 & 0 & 0 & 0 & 0 & 0 & 0 & 0 & 0 \\
    0 & 0 & 0 & 0 & 0 & 0 & 0 & 0 & 1 & 0 \\
    0 & 0 & 0 & 0 & 0 & 0 & 0 & 0 & 0 & 0 \\
    0 & 0 & 0 & 0 & 0 & 0 & 0 & 0 & 0 & 0 \\
    0 & 0 & 0 & 0 & 0 & 0 & 0 & 0 & 0 & 0 \\
    0 & 0 & 0 & 0 & 0 & 0 & 0 & 0 & 0 & 0 \\
    0 & 0 & 0 & 0 & 0 & 0 & 0 & 0 & 0 & 0 \\
    0 & 0 & 0 & 0 & 0 & 0 & 0 & 0 & 0 & 0 \\
    0 & -1 & 0 & 0 & 0 & 0 & 0 & 0 & 0 & 0 \\
    0 & 0 & 0 & 0 & 0 & 0 & 0 & 0 & 0 & 0
\end{array}
\right)
$$\\
$$
K=\left(
\begin{array}{cccccccccc}
    0 & 0 & 0 & 0 & 0 & 1 & 0 & 0 & 0 & 0 \\
    0 & 0 & 0 & 0 & 0 & 0 & 1 & 0 & 0 & 0 \\
    0 & 0 & 0 & 0 & 0 & 0 & 0 & 0 & 0 & 1 \\
    0 & 0 & 0 & 0 & 0 & 0 & 0 & 0 & 0 & 0 \\
    0 & 0 & 0 & 0 & 0 & 0 & 0 & 0 & 0 & 0 \\
    -1 & 0 & 0 & 0 & 0 & 0 & 0 & 0 & 0 & 0 \\
    0 & -1 & 0 & 0 & 0 & 0 & 0 & 0 & 0 & 0 \\
    0 & 0 & 0 & 0 & 0 & 0 & 0 & 0 & 0 & 0 \\
    0 & 0 & 0 & 0 & 0 & 0 & 0 & 0 & 0 & 0 \\
    0 & 0 & -1 & 0 & 0 & 0 & 0 & 0 & 0 & 0
\end{array}
\right)
$$\\
$$
L=\left(
\begin{array}{ccccccccccc}
    0 & 0 & 0 & 0 & 0 & 0 & 0 & 0 & 0 & 0 \\
    0 & 0 & 0 & 0 & 0 & 0 & 0 & 1 & 0 & 0 \\
    0 & 0 & 0 & 0 & 0 & 0 & 0 & 0 & 0 & 0 \\
    0 & 0 & 0 & 0 & 0 & 0 & 0 & 0 & 0 & 0 \\
    0 & 0 & 0 & 0 & 0 & 0 & 0 & 0 & 0 & 0 \\
    0 & 0 & 0 & 0 & 0 & 0 & 0 & 0 & 0 & 0 \\
    0 & 0 & 0 & 0 & 0 & 0 & 0 & 0 & 0 & 0 \\
    0 & -1 & 0 & 0 & 0 & 0 & 0 & 0 & 0 & 0 \\
    0 & 0 & 0 & 0 & 0 & 0 & 0 & 0 & 0 & 0 \\
    0 & 0 & 0 & 0 & 0 & 0 & 0 & 0 & 0 & 0
\end{array}
\right)
$$\\
$$
S(\Bbb{Z})=up+\frac{1}{2}(p^{xxx})^2+\frac
{\sigma}{2}w^2+u^3-p^xv-p^{xx}u-p^{xt}p-p^{xy}w.
$$

$\nabla S$ is the gradient of $S$ with respect to the standard
inner product on $\Bbb{R}^{10}$. The system (2.21) is a
Hamiltonian formulation of the KP equation on a multisymplectic
structure. Although this formulation is not the best, we give a
constructive method to get multisymplectic form which proposed by
Bridges [3]. It should be pay attention that (2.21) is different
with the multisymplectic form of KP equation in [4]. For equation
(2.21), there is a conservation law
\begin{equation}
\partial_t(d\Bbb{Z}\wedge Md\Bbb{Z})+\partial_x(d\Bbb{Z}\wedge
Kd\Bbb{Z})+\partial_y(d\Bbb{Z}\wedge Ld\Bbb{Z})=0.
\end{equation}
Substituting $M$,$K$,$L$ into (2.22) leads to
\begin{equation}
\begin{array}{l}
\frac {\partial}{\partial t}(d\varphi_{x}\wedge d\varphi_{xx})+\frac
{\partial}{\partial
x}(d\varphi\wedge(-6\varphi_{xxx}d\varphi_{xx}-6\varphi_{xx}d\varphi_{xxx}-\sigma d\varphi_{xyy}-2d\varphi_{xxt}-d\varphi_{xxxxx})\\
+d\varphi_x\wedge(6\varphi_{xx}d\varphi_{xx}+d\varphi_{xt}+d\varphi_{xxxx})+d\varphi_{xx}\wedge d(-\varphi_{xxx}))+\frac{\partial}{\partial
y}(d\varphi_x\wedge d(\sigma \varphi_{xy}))=0.
\end{array}
\end{equation}

This multisymplectic conservation law (2.22) consistent with our
theorem 2.3. We can regard the conservation law (2.23) as a
corollary of the theorem 2.3. Let $V$,$W$ be $\pi_{XY}$-vertical
and have the expressions
$V^\varphi\frac{\partial}{\partial\varphi}$,$W^\varphi\frac{\partial}{\partial\varphi}$.
Thus the corresponding $j^5(V)$ and $j^5(W)$ have the coordinate
expressions $(V^\varphi,
V^\varphi_x,V^\varphi_y,V^\varphi_t,V^\varphi_{xx},V^\varphi_{yy},V^\varphi_{tt},V^\varphi_{xt},V^\varphi_{xy},V^\varphi_{yt},
V^\varphi_{xxx}$,\newline
$V^\varphi_{xtx},V^\varphi_{xxy},V^\varphi_{xtt},V^\varphi_{xyy},V^\varphi_{ttt},V^\varphi_{yyy},V^\varphi_{tyy},V^\varphi_{tty},
V^\varphi_{xxxx},V^\varphi_{xxyy},V^\varphi_{xxtt},V^\varphi_{xxxy},V^\varphi_{xxxt},
V^\varphi_{xyyy},V^\varphi_{xttt},V^\varphi_{tttt}$,\newline $
V^\varphi_{yyyy},V^\varphi_{tyyy},V^\varphi_{ttyy},V^\varphi_{ttty},V^\varphi_{xxxxx},V^\varphi_{xxxxy},V^\varphi_{xxxxt},
V^\varphi_{xxxyy},V^\varphi_{xxxtt},V^\varphi_{xxyyy},V^\varphi_{xxttt},V^\varphi_{xyyyy},V^\varphi_{xtttt}$,\newline
$
V^\varphi_{yyyyy},V^\varphi_{ttttt},V^\varphi_{yyyyt},V^\varphi_{yyytt},V^\varphi_{yyttt},V^\varphi_{ytttt})$
and $(W^\varphi,
W^\varphi_x,W^\varphi_y,W^\varphi_t,W^\varphi_{xx},W^\varphi_{yy},W^\varphi_{tt},W^\varphi_{xt},W^\varphi_{xy}$,\newline
$W^\varphi_{yt},
 W^\varphi_{xxx},W^\varphi_{xtx},W^\varphi_{xxy},W^\varphi_{xtt},W^\varphi_{xyy},W^\varphi_{ttt},W^\varphi_{yyy},W^\varphi_{tyy},W^\varphi_{tty},
W^\varphi_{xxxx},W^\varphi_{xxyy},W^\varphi_{xxtt},W^\varphi_{xxxy}$,\newline
$W^\varphi_{xxxt},W^\varphi_{xyyy},W^\varphi_{xttt},W^\varphi_{tttt},
W^\varphi_{yyyy},W^\varphi_{tyyy},W^\varphi_{ttyy},W^\varphi_{ttty},W^\varphi_{xxxxx},W^\varphi_{xxxxy},W^\varphi_{xxxxt},
W^\varphi_{xxxyy}$,\newline
$W^\varphi_{xxxtt},W^\varphi_{xxyyy},W^\varphi_{xxttt},W^\varphi_{xyyyy},W^\varphi_{xtttt},
W^\varphi_{yyyyy},W^\varphi_{ttttt},W^\varphi_{yyyyt},W^\varphi_{yyytt},W^\varphi_{yyttt},W^\varphi_{ytttt}).$
\vbox{ \quad}

We can compute
$$
\begin{array}{rcl}
 & & j^5(\varphi)^\ast(j^5(V)\rfloor j^5(W)\rfloor\Omega_L)=
(W^\varphi_{xx}V^\varphi_x-W_x^\varphi V_{xx}^\varphi)dx\wedge
dy\\
& & +(\sigma W_x^\varphi V^\varphi_{xy}-\sigma W_{xy}^\varphi
V^\varphi_{x})dx\wedge dt+[W^\varphi_{xx}V^\varphi_{xxx}-W_{xxx}^\varphi
V_{xx}^\varphi\\
& & -W^\varphi_x(V_{xt}^\varphi+6\varphi_{xx}V^\varphi_{xx}+V^\varphi_{xxxx})+
V^\varphi_x(W_{xt}^\varphi+6\varphi_{xx}W^\varphi_{xx}+W^\varphi_{xxxx})
\\
& & +V^\varphi(-2W_{xxt}^\varphi-6\varphi_{xx}W^\varphi_{xxx}-6\varphi_{xxx}W^\varphi_{xx}-\sigma
W^\varphi_{xyy}-W^\varphi_{xxxxx})\\
& &
+W^\varphi(2V_{xxt}^\varphi+6\varphi_{xx}V^\varphi_{xxx}+6\varphi_{xxx}V^\varphi_{xx}+\sigma
V^\varphi_{xyy}+V^\varphi_{xxxxx})]dy\wedge dt.
\end{array}
$$

By the Stokes' formula, we can obtain
$$
\begin{array}{rcl}
 & &\int_{U}\frac{\partial}{\partial t}
(W^\varphi_{xx}V^\varphi_x-W_x^\varphi V_{xx}^\varphi)dx\wedge
dy\wedge dt-\frac{\partial}{\partial y}(\sigma W_x^\varphi
V^\varphi_{xy}-\sigma W_{xy}^\varphi
V^\varphi_{x})dx\wedge dy\wedge dt\\
& & +\frac{\partial}{\partial
x}[W^\varphi_{xx}V^\varphi_{xxx}-W_{xxx}^\varphi
V_{xx}^\varphi-W^\varphi_x(V_{xt}^\varphi+6\varphi_{xx}V^\varphi_{xx}+V^\varphi_{xxxx})+
V^\varphi_x(W_{xt}^\varphi+6\varphi_{xx}W^\varphi_{xx}\\
& &+W^\varphi_{xxxx})
+V^\varphi(-2W_{xxt}^\varphi-6\varphi_{xx}W^\varphi_{xxx}-6\varphi_{xxx}W^\varphi_{xx}-\sigma
W^\varphi_{xyy}-W^\varphi_{xxxxx})\\
& &
+W^\varphi(2V_{xxt}^\varphi+6\varphi_{xx}V^\varphi_{xxx}+6\varphi_{xxx}V^\varphi_{xx}+\sigma
V^\varphi_{xyy}+V^\varphi_{xxxxx})]dx\wedge dy\wedge dt=0.
\end{array}
$$
Since $U$ is arbitrary, we get
\begin{equation}
\begin{array}{l}
\frac{\partial}{\partial
t}(W^\varphi_{xx}V^\varphi_x-W_x^\varphi V_{xx}^\varphi)-\frac{\partial}{\partial y}(\sigma W_x^\varphi
V^\varphi_{xy}-\sigma W_{xy}^\varphi
V^\varphi_{x})
+\frac{\partial}{\partial
x}[W^\varphi_{xx}V^\varphi_{xxx}-\\
W_{xxx}^\varphi
V_{xx}^\varphi-W^\varphi_x(V_{xt}^\varphi+6\varphi_{xx}V^\varphi_{xx}+V^\varphi_{xxxx})+
V^\varphi_x(W_{xt}^\varphi+6\varphi_{xx}W^\varphi_{xx}+W^\varphi_{xxxx})\\
+V^\varphi(-2W_{xxt}^\varphi-6\varphi_{xx}W^\varphi_{xxx}-6\varphi_{xxx}W^\varphi_{xx}-\sigma
W^\varphi_{xyy}-W^\varphi_{xxxxx})\\
+W^\varphi(2V_{xxt}^\varphi+6\varphi_{xx}V^\varphi_{xxx}+6\varphi_{xxx}V^\varphi_{xx}+\sigma
V^\varphi_{xyy}+V^\varphi_{xxxxx})]=0.
\end{array}
\end{equation}
If we let
$$
M^t=\left(
\begin{array}{ccccccccccc}
    0 & 0 & 0 & 0 & 0 & 0 & 0 & 0 & 0 & 0 \\
    0 & 0 & 1 & 0 & 0 & 0 & 0 & 0 & 0 & 0 \\
    0 & -1 & 0 & 0 & 0 & 0 & 0 & 0 & 0 & 0 \\
    0 & 0 & 0 & 0 & 0 & 0 & 0 & 0 & 0 & 0 \\
    0 & 0 & 0 & 0 & 0 & 0 & 0 & 0 & 0 & 0 \\
    0 & 0 & 0 & 0 & 0 & 0 & 0 & 0 & 0 & 0 \\
    0 & 0 & 0 & 0 & 0 & 0 & 0 & 0 & 0 & 0 \\
    0 & 0 & 0 & 0 & 0 & 0 & 0 & 0 & 0 & 0 \\
    0 & 0 & 0 & 0 & 0 & 0 & 0 & 0 & 0 & 0 \\
    0 & 0 & 0 & 0 & 0 & 0 & 0 & 0 & 0 & 0
\end{array}
\right)
$$\\
$$
M^x=\left(
\begin{array}{cccccccccc}
    0 & 0 & -6\varphi_{xxx} & 0 & 0 & -6\varphi_{xx} & -\sigma & -2 & 0 & -1 \\
    0 & 0 & 6\varphi_{xx} & 0 & 1 & 0 & 0 & 0 & 1 & 0 \\
    6\varphi_{xxx} & -6\varphi_{xx} & 0 & 0 & 0 & -1 & 0 & 0 & 0 & 0 \\
    0 & 0 & 0 & 0 & 0 & 0 & 0 & 0 & 0 & 0 \\
    0 & -1 & 0 & 0 & 0 & 0 & 0 & 0 & 0 & 0 \\
    6\varphi_{xx} & 0 & 1 & 0 & 0 & 0 & 0 & 0 & 0 & 0 \\
    \sigma & 0 & 0 & 0 & 0 & 0 & 0 & 0 & 0 & 0 \\
    2 & 0 & 0 & 0 & 0 & 0 & 0 & 0 & 0 & 0 \\
    0 & -1 & 0 & 0 & 0 & 0 & 0 & 0 & 0 & 0 \\
    1 & 0 & 0 & 0 & 0 & 0 & 0 & 0 & 0 & 0
\end{array}
\right)
$$\\
$$
M^y=\left(
\begin{array}{cccccccccc}
    0 & 0 & 0 & 0 & 0 & 0 & 0 & 0 & 0 & 0 \\
    0 & 0 & 0 & \sigma & 0 & 0 & 0 & 0 & 0 & 0 \\
    0 & 0 & 0 & 0 & 0 & 0 & 0 & 0 & 0 & 0 \\
    0 & -\sigma & 0 & 0 & 0 & 0 & 0 & 0 & 0 & 0 \\
    0 & 0 & 0 & 0 & 0 & 0 & 0 & 0 & 0 & 0 \\
    0 & 0 & 0 & 0 & 0 & 0 & 0 & 0 & 0 & 0 \\
    0 & 0 & 0 & 0 & 0 & 0 & 0 & 0 & 0 & 0 \\
    0 & 0 & 0 & 0 & 0 & 0 & 0 & 0 & 0 & 0 \\
    0 & 0 & 0 & 0 & 0 & 0 & 0 & 0 & 0 & 0 \\
    0 & 0 & 0 & 0 & 0 & 0 & 0 & 0 & 0 & 0
\end{array}
\right)
$$\\
\begin{equation}
M^t(x,y)=x^T M^ty, M^x(x,y)=x^T M^xy, M^y(x,y)=x^T M^y y
\end{equation}
and set
$$
j^5(V)=(V^\varphi, V^\varphi_x,V^\varphi_{xx},V^\varphi_{xy},V^\varphi_{xt},V^\varphi_{xxx},V^\varphi_{xyy},V^\varphi_{xxt},V^\varphi_{xxxx},V^\varphi_{xxxxx})
$$
$$
j^5(W)=(W^\varphi, W^\varphi_x,W^\varphi_{xx},W^\varphi_{xy},W^\varphi_{xt},W^\varphi_{xxx},W^\varphi_{xyy},W^\varphi_{xxt},W^\varphi_{xxxx},W^\varphi_{xxxxx})
$$
the others coordinates vanished. Then conservation law (2.24) can
be written as
\begin{equation}
\begin{array}{l}
 \frac
{\partial}{\partial t} M^t(j^5(V),j^5(W))+\frac
{\partial}{\partial x} M^x(j^5(V),j^5(W))+\frac
{\partial}{\partial y} M^y(j^5(V),j^5(W))=0\\
\end{array}
\end{equation}

Also since the translation invariance of the KP equation, we
choose $V=W=\varphi$ and take it into (2.26), conservation law
(2.26) becomes:
\begin{equation}
\begin{array}{l}
\frac {\partial}{\partial t}(d\varphi_{x}\wedge d\varphi_{xx})+\frac
{\partial}{\partial
x}(d\varphi\wedge(-6\varphi_{xxx}d\varphi_{xx}-6\varphi_{xx}d\varphi_{xxx}-\sigma d\varphi_{xyy}-2d\varphi_{xxt}-d\varphi_{xxxxx})\\
+d\varphi_x\wedge(6\varphi_{xx}d\varphi_{xx}+d\varphi_{xt}+d\varphi_{xxxx})+d\varphi_{xx}\wedge d(-\varphi_{xxx}))+\frac{\partial}{\partial
y}(d\varphi_x\wedge d(\sigma \varphi_{xy}))=0.
\end{array}
\end{equation}
i.e. the conservation law (2.23).

In the numerical study, the multisymplectic conservation law can
be used to design multisymplectic numerical schemes, i.e.,
numerical schemes which can preserve the multisymplectic
conservation law.

\section{ Multisymplectic Preissman Scheme for the KP Equation}
\setcounter{equation}{0}

In this section, we consider the multisymplectic Preissman scheme
for the KP equation. The equation (2.21) can be reformulated as

\begin{equation}
\left\{
\begin{array}{l}
\frac {\partial p^x}{\partial x}=0,\\
\frac {\partial
p^{xx}}{\partial x}+\frac {\partial p^{xy}}{\partial y}+\frac
{\partial p^{xt}}{\partial t}=-p^x,\\
\frac {\partial p^{xxx}}{\partial x}=p+3u^2-p^{xx},\\
\sigma
w-p^{xy}=0,\\
 u-p^{xt}=0,\\
\frac {\partial \varphi}{\partial x}=v,\\
\frac {\partial
v}{\partial x}=u,\\
\frac {\partial v}{\partial
y}=w,\\
\frac {\partial v}{\partial t}=p,\\
\frac {\partial
u}{\partial x}=p^{xxx}
\end{array}
\right.
\end{equation}

For convenience, we assumed that the spacing of the grid points in the $x,y,t$ directions is uniform respectively.
We apply the implicit midpoint discretization in time and in
space to (3.1), and obtain that:
\begin{equation}
\left\{
\begin{array}{l}
\frac{p^x_{i+1,j+\frac{1}{2},k+\frac{1}{2}}-p^x_{i,j+\frac{1}{2},
k+\frac{1}{2}}}{\bigtriangleup x}=0,\\
\frac{p^{xx}_{i+1,j+\frac{1}{2},k+\frac{1}{2}}-p^{xx}_{i,j+\frac{1}{2},
k+\frac{1}{2}}}{\bigtriangleup
x}+\frac{p^{xy}_{i+\frac{1}{2},j+1,k+\frac{1}{2}}-p^{xy}_{i+\frac{1}{2},j,
k+\frac{1}{2}}}{\bigtriangleup
y}\\
\quad \quad \quad \quad
+\frac{p^{xt}_{i+\frac{1}{2},j+\frac{1}{2},k+1}-p^{xt}_{i+\frac{1}{2},j+\frac{1}{2},
k}}{\bigtriangleup
t}=-p^x_{i+\frac{1}{2},j+\frac{1}{2},k+\frac{1}{2}},\\
\frac{p^{xxx}_{i+1,j+\frac{1}{2},k+\frac{1}{2}}-p^{xxx}_{i,j+\frac{1}{2},
k+\frac{1}{2}}}{\bigtriangleup
x}=p_{i+\frac{1}{2},j+\frac{1}{2},k+\frac{1}{2}}+3(u_{i+\frac{1}{2},j+\frac{1}{2},k+\frac{1}{2}})^2-p_{i+\frac{1}{2},j+\frac{1}{2},k+\frac{1}{2}}^{xx}\\
\sigma
w_{i+\frac{1}{2},j+\frac{1}{2},k+\frac{1}{2}}=p^{xy}_{i+\frac{1}{2},j+\frac{1}{2},k+\frac{1}{2}},\\
u_{i+\frac{1}{2},j+\frac{1}{2},k+\frac{1}{2}}=p^{xt}_{i+\frac{1}{2},j+\frac{1}{2},k+\frac{1}{2}},\\
\frac
{\varphi_{i+1,j+\frac{1}{2},k+\frac{1}{2}}-\varphi_{i,j+\frac{1}{2},k+\frac{1}{2}}}{\bigtriangleup
x}=v_{i+\frac{1}{2},j+\frac{1}{2},k+\frac{1}{2}},\\
\frac
{v_{i+1,j+\frac{1}{2},k+\frac{1}{2}}-v_{i,j+\frac{1}{2},k+\frac{1}{2}}}{\bigtriangleup
x}=u_{i+\frac{1}{2},j+\frac{1}{2},k+\frac{1}{2}},\\
\frac
{v_{i+\frac{1}{2},j+1,k+\frac{1}{2}}-v_{i+\frac{1}{2},j,k+\frac{1}{2}}}{\bigtriangleup
y}=w_{i+\frac{1}{2},j+\frac{1}{2},k+\frac{1}{2}},\\
\frac
{v_{i+\frac{1}{2},j+\frac{1}{2},k+1}-v_{i+\frac{1}{2},j+\frac{1}{2},k}}{\bigtriangleup
t}=p_{i+\frac{1}{2},j+\frac{1}{2},k+\frac{1}{2}},\\
\frac
{u_{i+1,j+\frac{1}{2},k+\frac{1}{2}}-u_{i,j+\frac{1}{2},k+\frac{1}{2}}}{\bigtriangleup
x}=p^{xxx}_{i+\frac{1}{2},j+\frac{1}{2},k+\frac{1}{2}},
\end{array}
\right.
\end{equation}
Where $\triangle x$ is the $x$-direction step, $\triangle y$ is the $y$-direction step, $\triangle t$ is the time step and
$u_{i+\frac{1}{2},j+\frac{1}{2},k+\frac{1}{2}}=u(i\triangle x+\frac{\triangle x}{2},j\triangle y+\frac{\triangle y}{2},k\triangle t+\frac{\triangle t}{2})$,
the others is similar.

In fact, the discretization result lead to the pressman scheme
\begin{equation}
\begin{array}{l}
\frac{1}{\bigtriangleup
t}M(\Bbb{Z}_{i+\frac{1}{2},j+\frac{1}{2},k+1}-\Bbb{Z}_{i+\frac{1}{2},j+\frac{1}{2},k})+\frac{1}{\bigtriangleup
x}K(\Bbb{Z}_{i+1,j+\frac{1}{2},k+\frac{1}{2}}-\Bbb{Z}_{i,j+\frac{1}{2},k+\frac{1}{2}})\\
+\frac{1}{\bigtriangleup
y}L(\Bbb{Z}_{i+\frac{1}{2},j+1,k+\frac{1}{2}}-\Bbb{Z}_{i+\frac{1}{2},j,k+\frac{1}{2}})=
\bigtriangledown S (\Bbb{Z}_{i+\frac{1}{2},j+\frac{1}{2},k+\frac{1}{2}})
\end{array}
\end{equation}
The (3.3) preserve the discrete multisymplectic conservation
law:
\begin{equation}
\begin{array}{l}
\frac {dv_{i+\frac{1}{2},j+\frac{1}{2},k+1}\wedge
dp^{xt}_{i+\frac{1}{2},j+\frac{1}{2},k+1}-dv_{i+\frac{1}{2},j+\frac{1}{2},k}\wedge
dp^{xt}_{i+\frac{1}{2},j+\frac{1}{2},k}}{\bigtriangleup t}+\\
\quad \quad  \frac {dv_{i+\frac{1}{2},j+1,k+\frac{1}{2}}\wedge
dp^{xy}_{i+\frac{1}{2},j+1,k+\frac{1}{2}}-dv_{i+\frac{1}{2},j,k+\frac{1}{2}}\wedge
dp^{xy}_{i+\frac{1}{2},j,k+\frac{1}{2}}}{\bigtriangleup y}+\\
\quad \quad  \frac
{d\varphi_{i+1,j+\frac{1}{2},k+\frac{1}{2}}\wedge
dp^{x}_{i+1,j+\frac{1}{2},k+\frac{1}{2}}-d\varphi_{i,j+\frac{1}{2},k+\frac{1}{2}}\wedge
dp^{x}_{i,j+\frac{1}{2},k+\frac{1}{2}}}{\bigtriangleup x}+\\
\quad \quad  \frac {dv_{i+1,j+\frac{1}{2},k+\frac{1}{2}}\wedge
dp^{xx}_{i+1,j+\frac{1}{2},k+\frac{1}{2}}-dv_{i,j+\frac{1}{2},k+\frac{1}{2}}\wedge
dp^{xx}_{i,j+\frac{1}{2},k+\frac{1}{2}}}{\bigtriangleup x}+\\
\quad \quad  \frac {du_{i+1,j+\frac{1}{2},k+\frac{1}{2}}\wedge
dp^{xxx}_{i+1,j+\frac{1}{2},k+\frac{1}{2}}-du_{i,j+\frac{1}{2},k+\frac{1}{2}}\wedge
dp^{xxx}_{i,j+\frac{1}{2},k+\frac{1}{2}}}{\bigtriangleup x}=0
\end{array}
\end{equation}
Although the Preissman scheme (3.3) is multisymplectic, it took too
efforts to realize. Hence we elimate the auxiliary variables
$\varphi, v,w,p,p^{x},p^{xx},p^{xy},p^{xt},p^{xxx}$ by a trivial
computation and obtain the following multisymplectic forty-five
points scheme:
\begin{equation}
\begin{array}{l}
\frac{1}{2 \bigtriangleup x \bigtriangleup t}\delta_y^2 \Delta_t^0
\{u_{i+2,j}^k+2u_{i+1,j}^k-2u_{i-1,j}^k+u_{i-2,j}^k\}\\
\quad \quad
+\frac{1}{\bigtriangleup x^4}\{\delta_y^2 \delta_t^2(u^k_{i+2,j}-4u_{i+1,j}^k+6u^k_{i,j}-4u_{i-1,j}^k+u_{i-2,j}^k)\}\\
\quad \quad +\frac {\sigma}{4\bigtriangleup
y^2}\{\delta_{t}^2\Delta
_y^2(u^k_{i+2,j}+4u_{i+1,j}^k+6u_{i,j}^k+4u_{i-1,j}^k+u_{i,j}^k)\}\\
\quad \quad +\frac{2}{\bigtriangleup x^2}(\overline{\delta}
f_{i,j+\frac{1}{2}}^{k+\frac{1}{2}}+\overline{\delta}
f_{i,j-\frac{1}{2}}^{k+\frac{1}{2}}+\overline{\delta}
f_{i,j+\frac{1}{2}}^{k-\frac{1}{2}}+\overline{\delta}
f_{i,j-\frac{1}{2}}^{k-\frac{1}{2}})=0
\end{array}
\end{equation}
where we denote $u_{i,j}^k=u(i\triangle x, j\triangle y,k\triangle t)$, $f_{i,j}^k=3(u(i\triangle x, j\triangle y,k\triangle t))^2$ and
\begin{equation}
\left\{
\begin{array}{l}
\Delta_t^0 u_{i,j}^k=u_{i,j}^{k+1}-u_{i,j}^{k-1}, \delta_y^2
u_{i,j}^k=u_{i,j+1}^k+2u_{i,j}^k+u_{i,j-1}^k, \\
\delta_t^2 u_{i,j}^k=u_{i,j}^{k+1}+2u_{i,j}^k+u_{i,j}^{k-1},
\Delta^2_y
u_{i,j}^k=u_{i,j+1}^{k}-2u_{i,j}^k+u_{i,j-1}^{k},\\
\overline{\delta}
f_{i,j}^k=f_{i+\frac{3}{2},j}^k-f_{i+\frac{1}{2},j}^k-f_{i-\frac{1}{2},j}^k+f_{i-\frac{3}{2},j}^k.
\end{array}
\right.
\end{equation}
\section{Some Numerical Results on Soliton and Solitary Waves}
\setcounter{equation}{0}

In this section, we test the forty-five points scheme on soliton and solitary
waves over long time intervals.
\begin{figure}[htbp]
\centerline {\hbox {\psfig{figure=
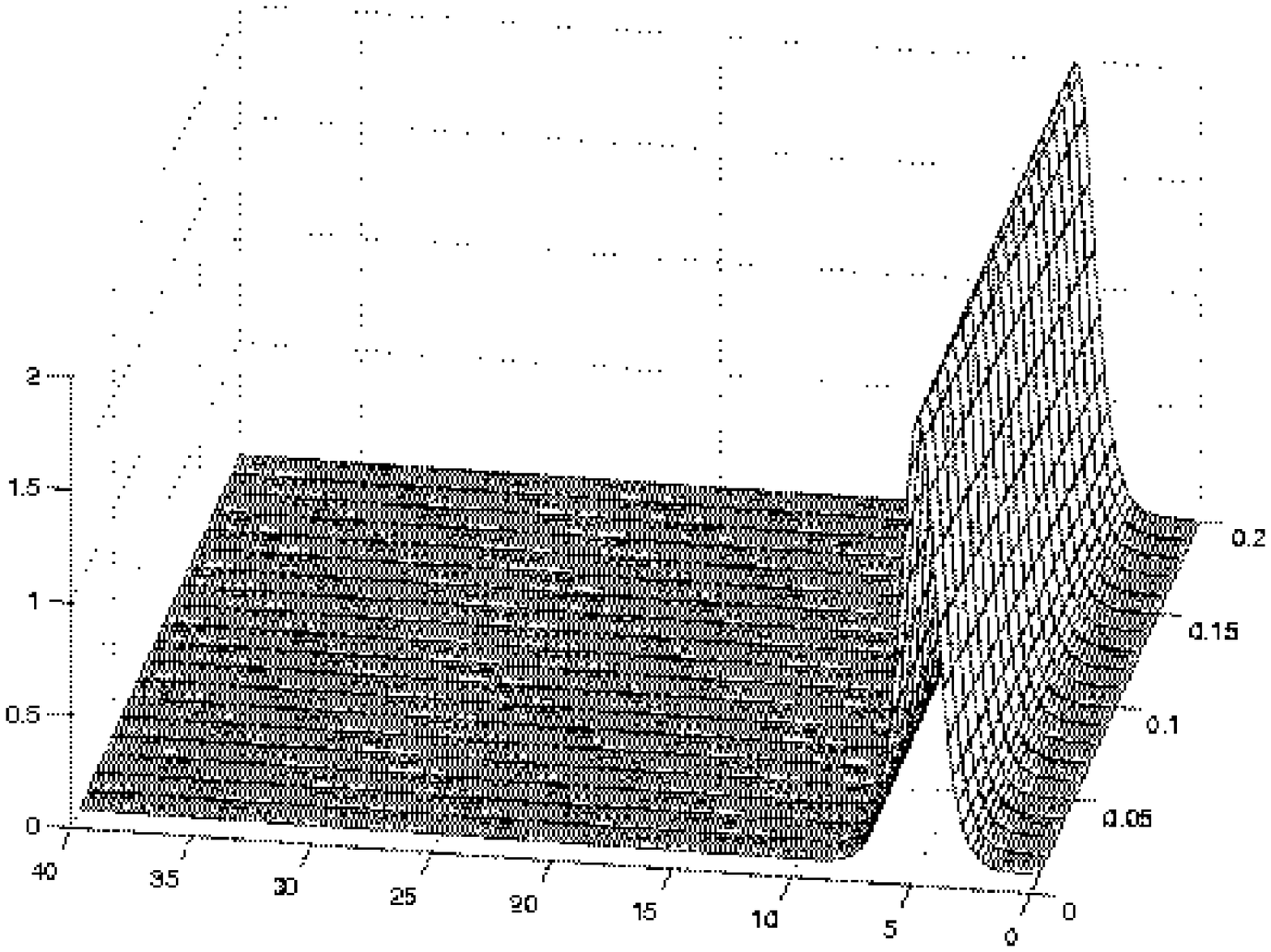,width=6.5cm,height=5.5cm} } \hskip 1cm \hbox {\psfig
{figure= 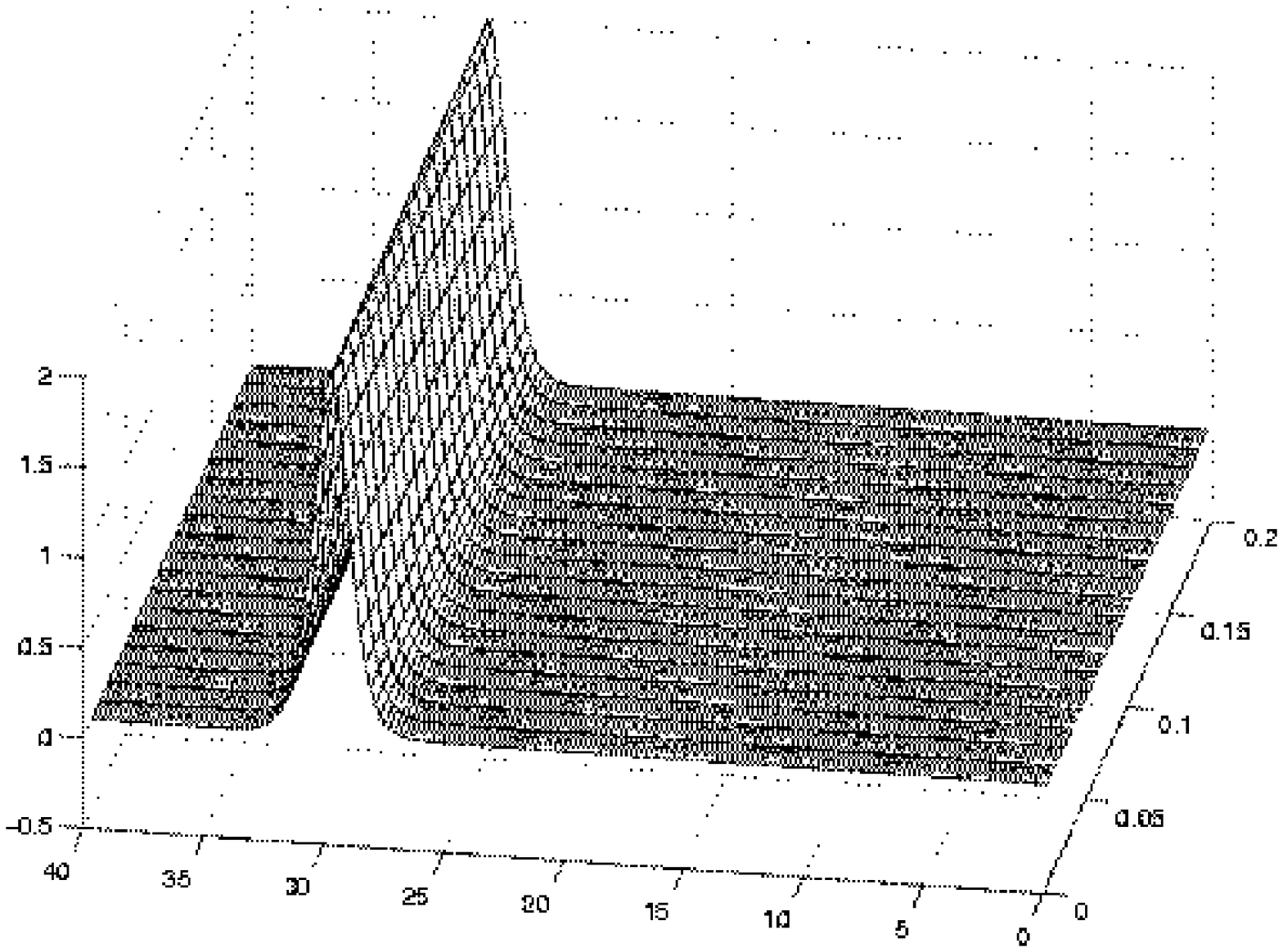,width=6.5cm,height=5.5cm} } }
\end{figure}

\quad Fig.1: one soliton at t=0, \hskip 3cm Fig.2: one soliton at
t=10.

At first, one line soliton of KPI equation is considered. We
choose small interval on $y$-direction just for computing
convenience, it have nothing with the scheme and result. We take
the following initial conditions
\begin{equation}
u(x,y,0)=2sech^2(x-\frac{\sqrt{2}}{2}y-6)
\end{equation}
and the exact boundary condition. The KPI equation has the
theoretic solution {}
$$
u(x,y,t)=2sech^2(x-\frac{\sqrt{2}}{2}y-\frac{5}{2}t-6)
$$
which represents one line-soliton propagating with the velocity
$\frac{5}{2}$ in the direction with the angle of
$tan^{-1}(\sqrt{2})$ to the positive $x$-axis. We take the test on
the domain $[0,40] \times [0,2]$ and choose $\bigtriangleup
x=0.2$,$\bigtriangleup y=0.1$,$\bigtriangleup t=0.01$. Figure 1
shows the initial condition and Figure 2 shows the numerical
solution at time $t=10 $ . In fact, it propagating 25-unit distant
as we indicated.

Next we try two line-soliton interaction.

We take the initial condition
\be
u(x,y,0)=2\sum^{2}_{i=1}k_i^2sech^2[k_i(x+\lambda_i y-x_{0 ,i})]
\ee
where
$k_1=1.0$,$k_2=1/\sqrt{2}$,$\lambda_1=-1/{\sqrt{3}}$,$\lambda_2=-1.0$
and $x_{0.1}=6.0$,$x_{0.2}=11.0$. and the exact boundary
condition.
The initial condition (4.2) corresponds to two line-solitons, each with amplitude $2k_i^2$ placed initially at $x=x_{0,i}$
and moving with velocity $v_i=4k_i^2-3\lambda_i^2$ along the $x$-axis ($i$=1,2).
\begin{figure}[htbp]
\centerline{\hbox{\psfig{figure=
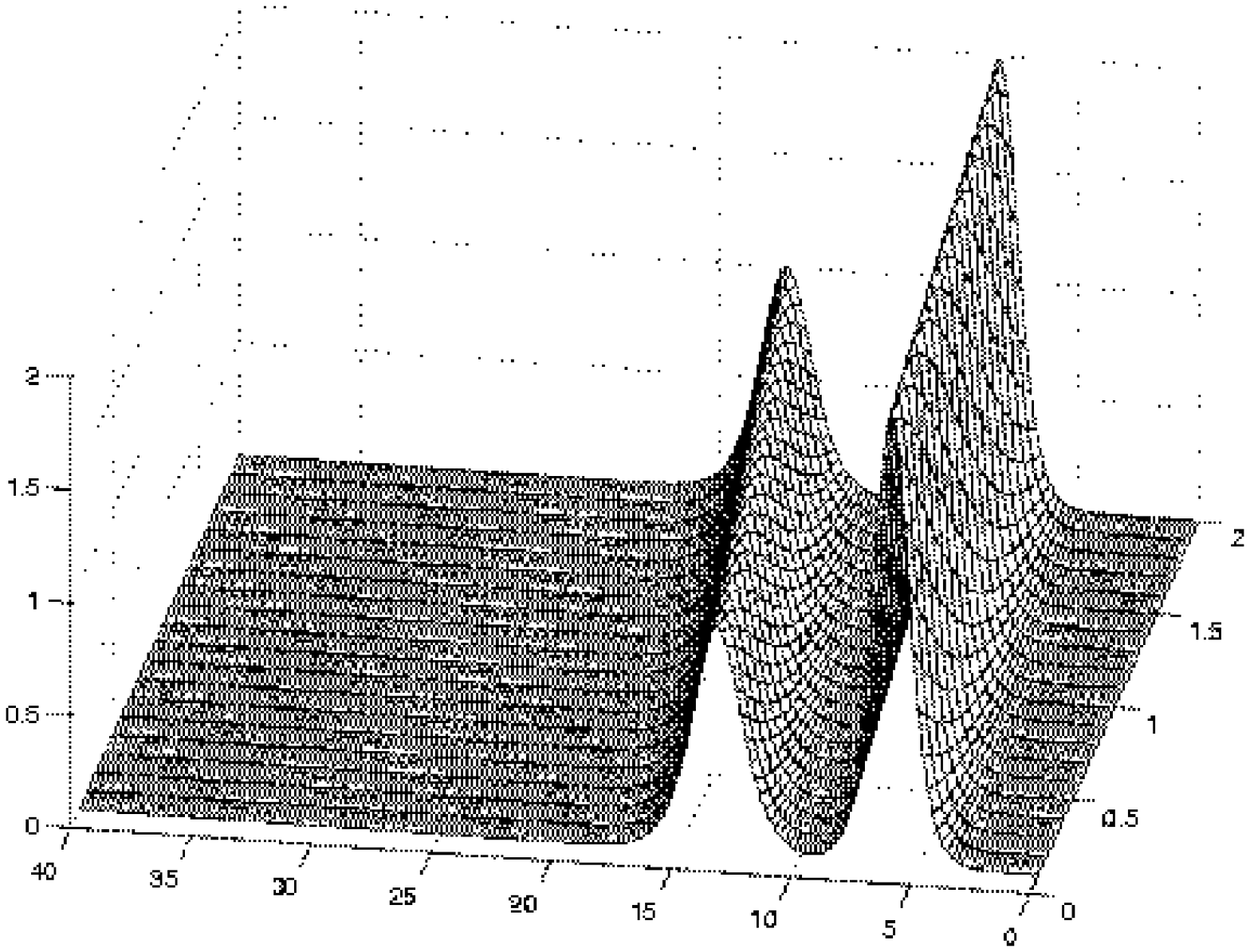,width=6.5cm,height=5.5cm}
}
\hskip 1cm
\hbox{\psfig{figure=
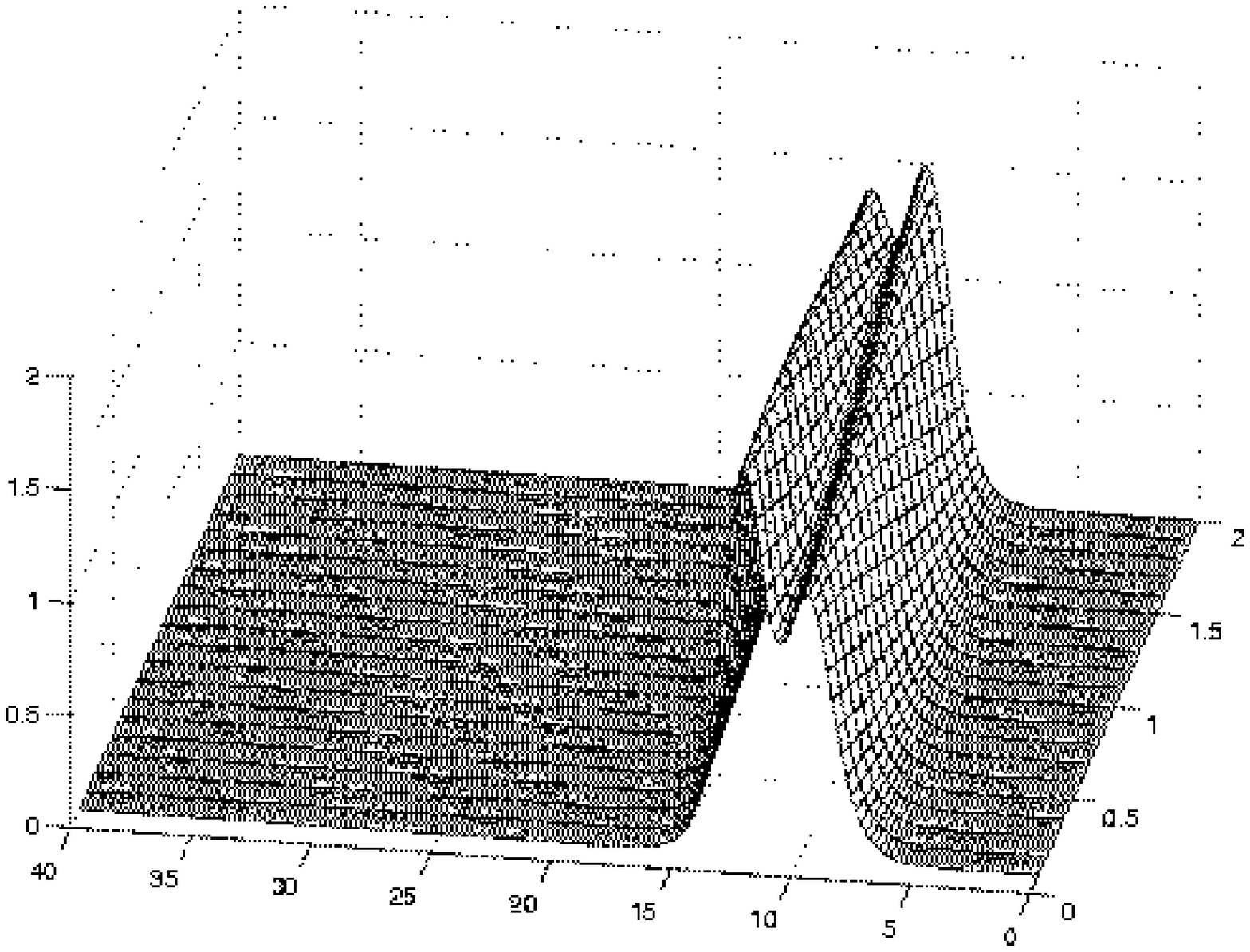,width=6.5cm,height=5.5cm}
}}
\end{figure}

\quad Fig.3: soliton interact at t=0, \hskip 2.5cm Fig.4: soliton
interact at t=1.5, \vskip 0.5cm

We carried out the computation on the domain $[0,40] \times [0,2]$, and choose $\bigtriangleup x=0.2$,$\bigtriangleup y=0.1$,$\bigtriangleup t=0.01$.
The initial condition (4.2) is showed in Figure 3. The larger line-soliton on the right will moves with a velocity 3.0 to the positive $x$-direction and
the smaller one on the left will moves with a velocity 1.0 to inverse direction. As time go on, they will collide with each other, it is show in Figure 4.
Figure 5 shows that the two line-solitons have separated completely after collide and restored their original shape by the $t=3$.

\begin{figure}[htbp]
\centerline{\hbox{\psfig{figure=
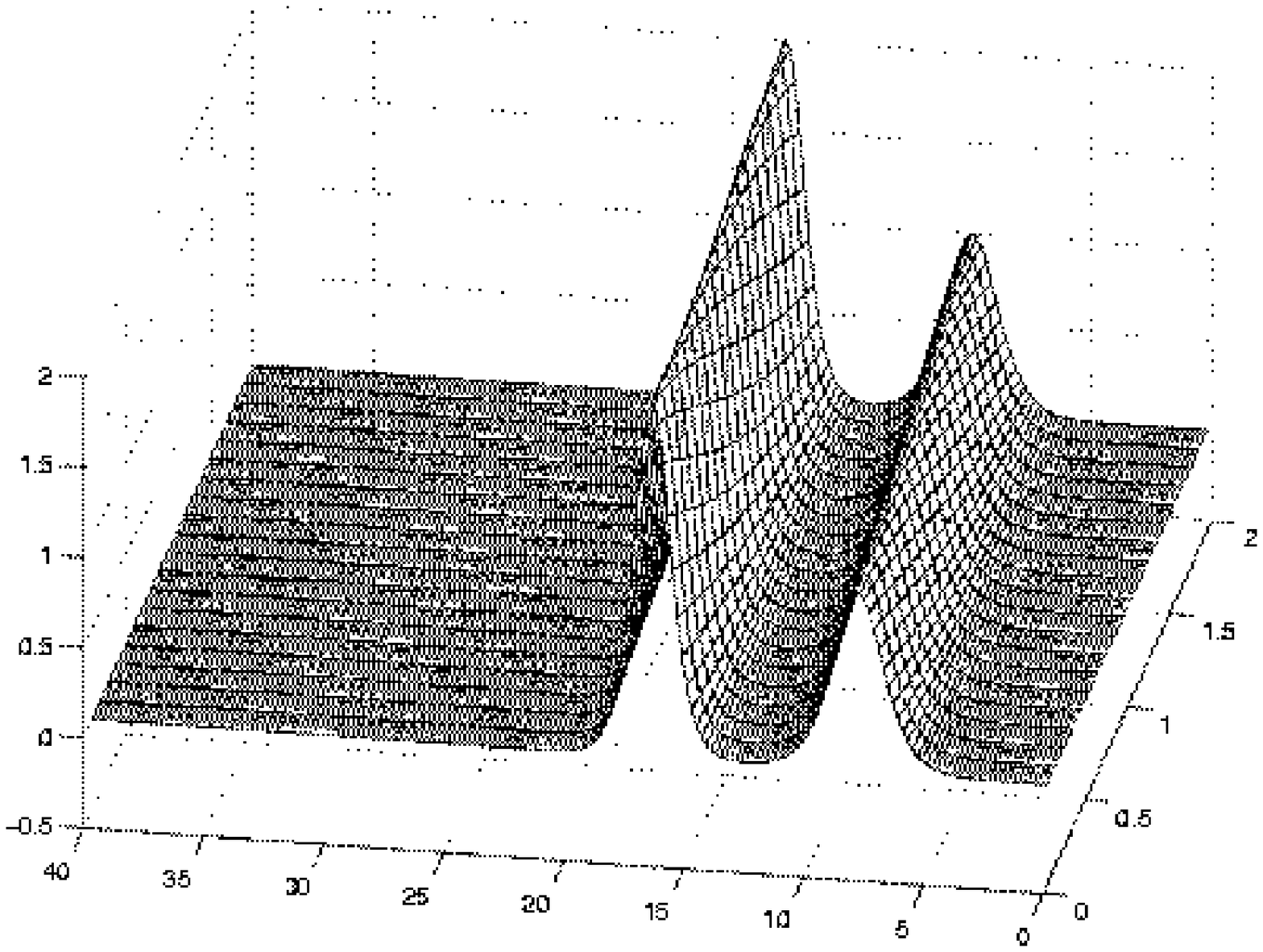,width=6.5cm,height=5.5cm}
}}
\end{figure}

\hskip 4cm Fig.5: soliton interact at t=3.

We also consider the lump type solitary waves of the KPI equation.
The lump type initial condition used for test is \be
u(x,y,0)=4\frac{(-(x-x_0)^2+\mu^2(y-y_0)^2+\frac{1}{\mu^2})
}{((x-x_0)^2+\mu^2(y-y_0)^2+\frac{1}{\mu^2}) }, \ee where the
parameters $\mu^2$=1.0, $x_0$=10.0, $y_0$=10.0, and we adopt the
exact boundary condition.
\begin{figure}[htbp]
\centerline {\hbox {\psfig{figure=
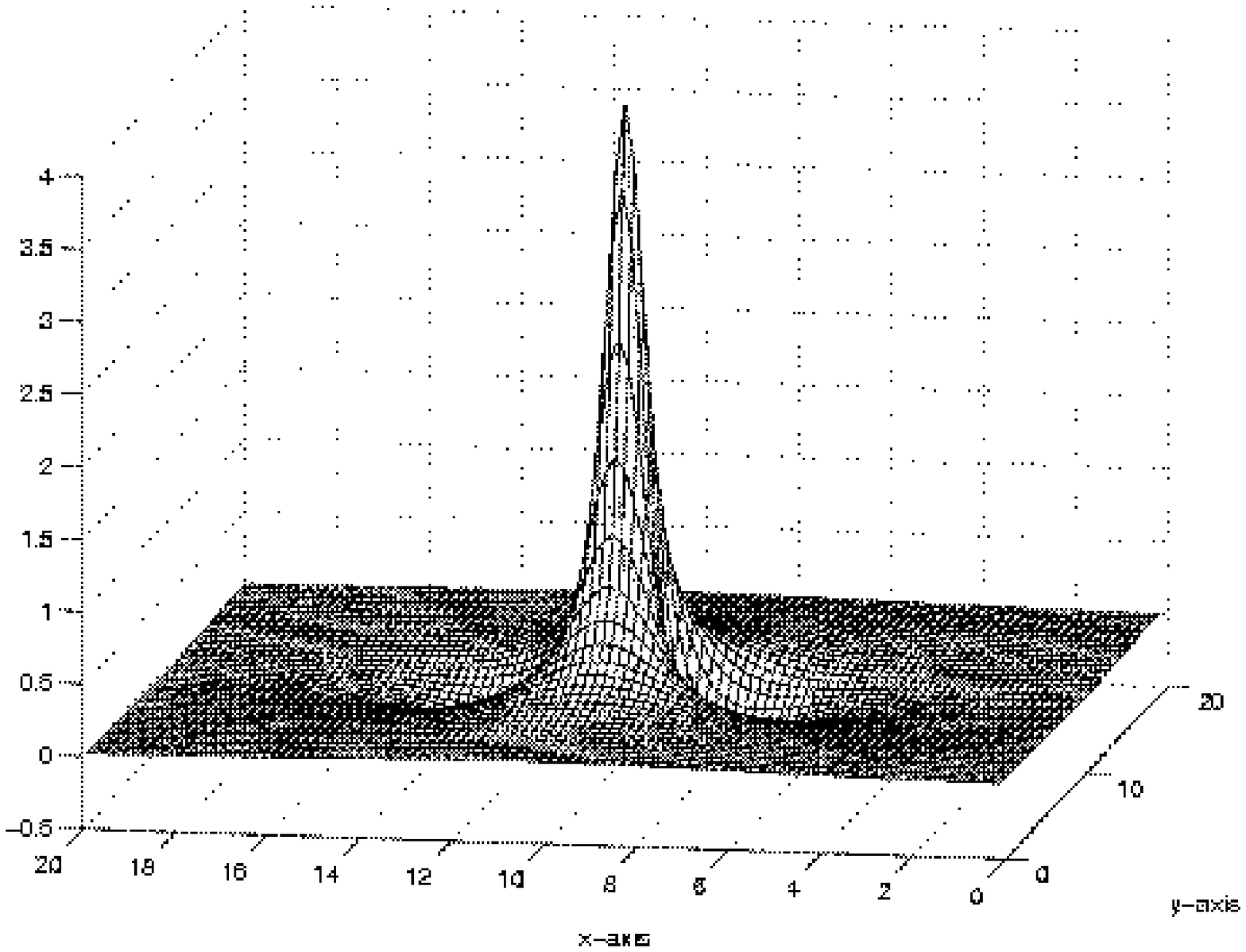,width=6.5cm,height=5.5cm} } \hskip 1cm \hbox {\psfig
{figure= 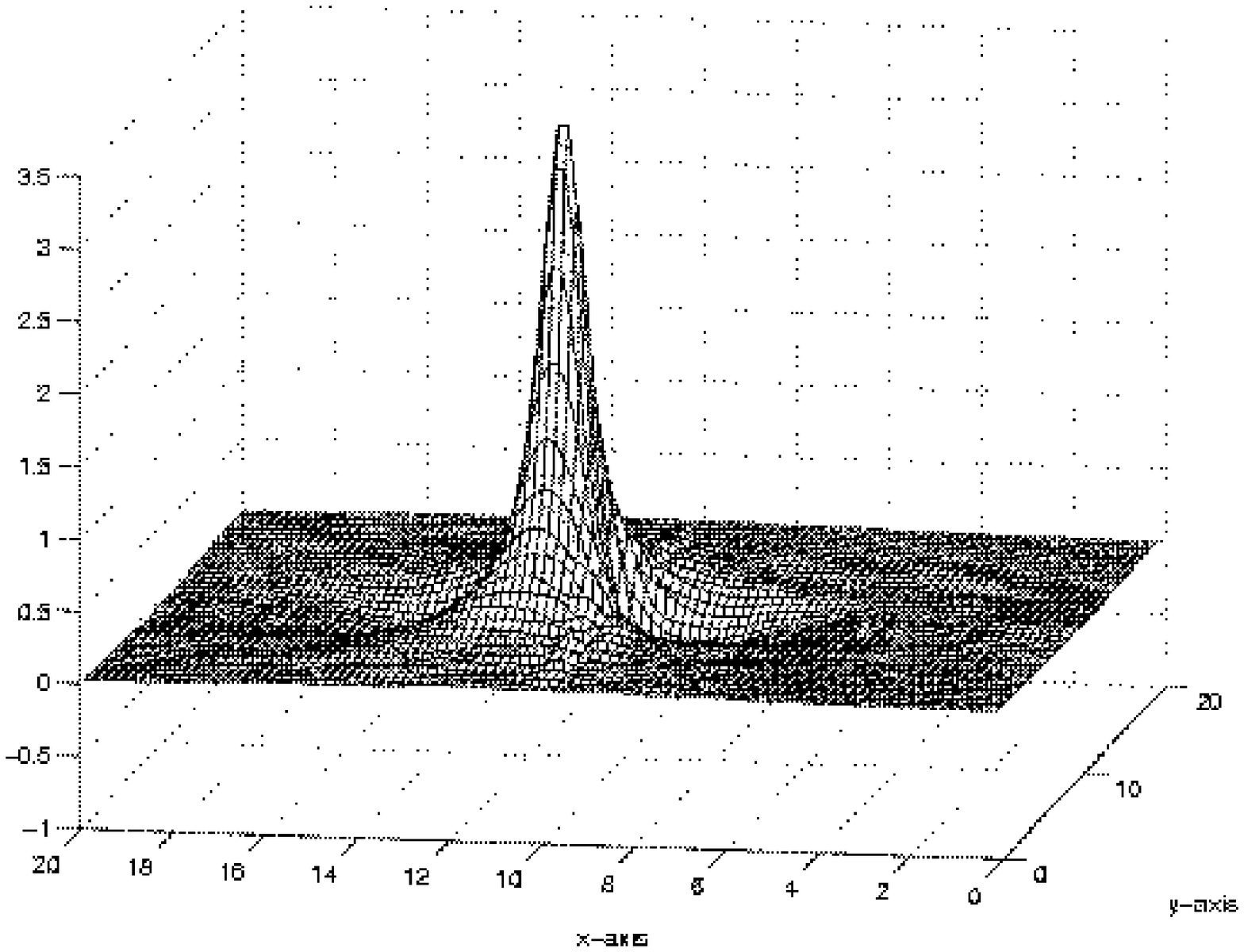,width=6.5cm,height=5.5cm} } }
\end{figure}

Fig.6: lump type solitary wave at t=0, \hskip 1cm Fig.7: lump type
solitary wave at t=0.5. \vskip 0.5cm

We compute in a rectangle $[0,20]\times[0,20]$ and choose
$\bigtriangleup x=0.1$,$\bigtriangleup y=0.2$,$\bigtriangleup
t=0.01$. Figure 6 shows the initial condition and figure 7 shows
the numerical solution at time $t=0.5$. By the time $t=1$, the
result is showed in figure 8. The lump solution of the KPI
equation can be expressed as \be
u(x,y,t)=4\frac{(-(x-x_0-3\mu^2t)^2+\mu^2(y-y_0)^2+\frac{1}{\mu^2})
}{((x-x_0-3\mu^2t)^2+\mu^2(y-y_0)^2+\frac{1}{\mu^2}) }. \ee
According to (4.4), this lump type solitary wave will move to the
positive $x$-direction with velocity $3\mu^2$. We can see the
moving of the lump solitary wave from the graph.
\begin{figure}[htbp]
\centerline{\hbox{\psfig{figure=
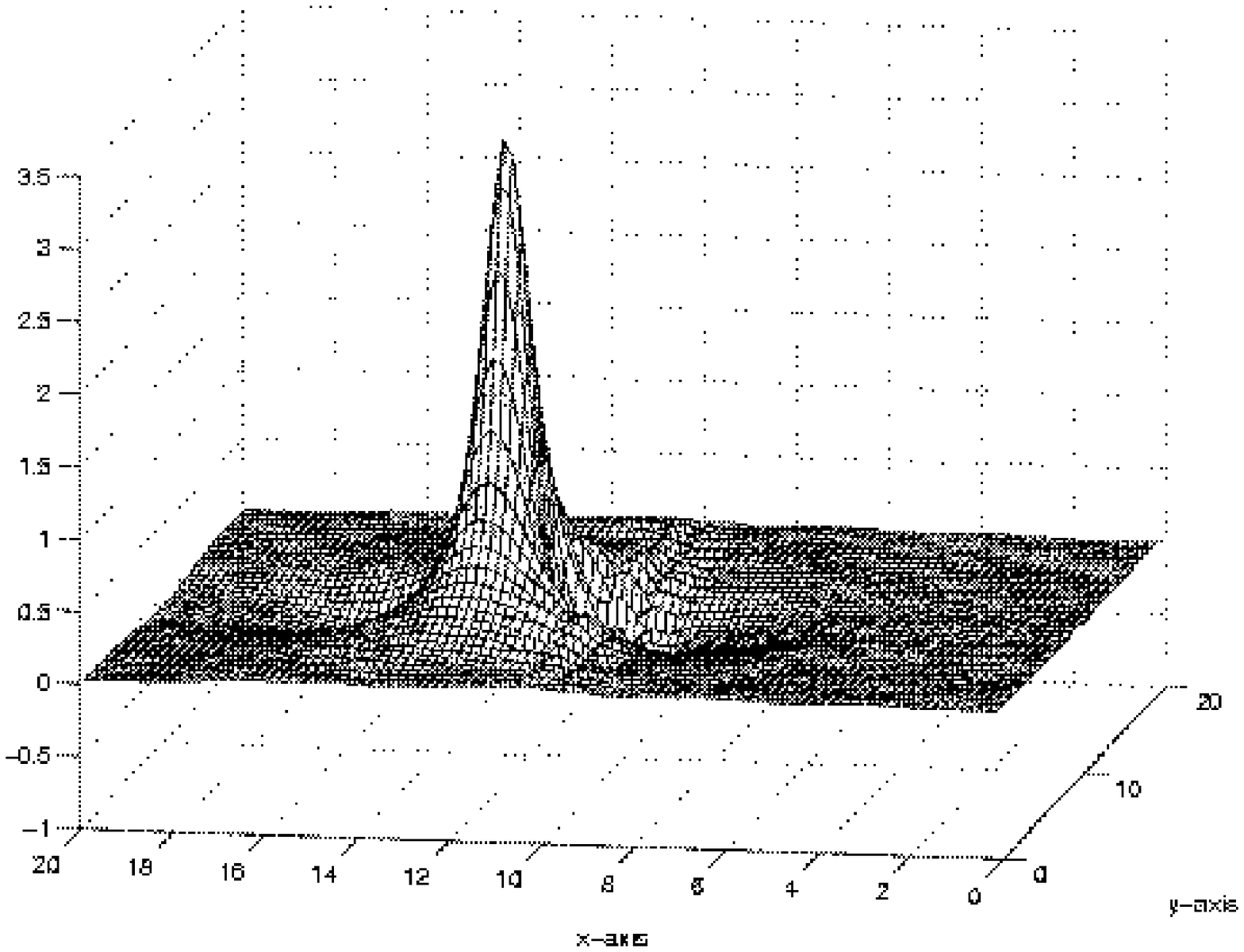,width=6.5cm,height=5.5cm}
}}
\end{figure}

\hskip 4cm Fig.8: lump type solitary wave at t=1. \vskip 0.5cm

A future task is expected to find a proper numerical boundary
conditions for collision of the two lump type solitary waves.


\end{document}